# Identifying fatigue crack initiation through analytical calculation of temporal compliance calibrated with Computed Tomography

Ritam Pal[a], Amrita Basak[a,*]

[a] Department of Mechanical Engineering, The Pennsylvania State University, University Park, PA 16802, USA

[*] Communicating author: aub1526@psu.edu

## Abstract

Fatigue failure is ubiquitous in engineering applications. While the total fatigue life is critical to understanding a component's operational life, for safety, regulatory compliance, and predictive maintenance, the characterization of initiation life is important. Traditionally, initiation life is characterized by potential drop method, acoustic emission technique, and strain-based measurements. However, the primary challenge with these methods lies in the necessity of calibration for each new material system. The difficulties become even more aggravated for additively manufactured components, where fatigue properties are reported to vary widely in the open literature. In this work, an analytical methodology is utilized to evaluate the initiation life of two different materials such as AlSi10Mg and SS316L, fabricated via laser-powder bed fusion (L-PBF) technique. The processing parameters are selected such that AlSi10Mg behaves like a brittle material while SS316L shows ductile behavior. A custom fatigue testing apparatus is used inside Computed Tomography (CT) for evaluating fatigue initiation. The apparatus reports load-displacement data, which is post-processed using an analytical approach to calculate the evolution of material compliance. The results indicate that crack initiation during fatigue loading is marked by a noticeable change in compliance. The analytical technique shows a maximum difference of 4.8% in predicting initiation life compared to CT imaging. These findings suggest that compliance monitoring can effectively identify fatigue initiation in various materials.

## 1. Introduction

Additive manufacturing (AM) such as laser-powder bed fusion (L-PBF) is widely used to fabricate parts with complex and intricate geometries [1]. L-PBF consolidates powder feedstock into a desired shape and geometry using a high-power laser beam [2]. The process parameters influence the microstructures of the components [3], which in turn affect the mechanical properties, especially fatigue [4]. Fatigue failure consists of three stages: crack initiation, propagation, and rapid failure [5]. Over each cycle, strain localizations occur, and they accumulate during fatigue loading, leading to crack initiation. The process of crack initiation involves the formation of two new surfaces, minimizing the accumulated strain energy. After crack initiation, the crack tip propagates with each loading cycle, decreasing the load-bearing area as the crack progresses. Finally, rapid failure occurs when the specimen is overstressed, and the rapid failure zone is devoid of any crack propagation marks. Depending on material properties, differences in fracture surfaces may be observed. For example, in brittle materials, the crack propagation zone is negligible on the fractured surface as the material does not undergo significant plastic deformation between crack initiation and final failure. On the contrary, in ductile materials, the crack tip undergoes plastic deformation with a visible propagation zone. The propagation period of a crack is also dependent on whether it is a short or a long fatigue crack [6].

Crack propagation in traditionally fabricated materials is well-established [7], [8]. However, crack initiation is an ongoing topic of investigation in the field of fatigue damage [9], [10], [11], [12]. Fatigue investigations have been carried out using acoustic emission techniques [13], potential drop methods [14], ultrasonic methods [15], infrared thermography [16], and strain-based techniques [17] to monitor crack initiation in different specimens. In the early years, the potential drop technique was mostly used to obtain the initiation life of specimens [18], [19]. The technique was based on the change in potential voltage across the specimen while the specimens underwent fatigue loading. With the advent of non-destructive testing technology, acoustic emission techniques gained the attention of researchers [20]. From acoustic emission signals, acoustic information entropy and acoustic signal amplitude were extracted to evaluate fatigue initiation in a specimen [21], [22]. Bjorheim et al. [23] provided an extensive overview of the measurement procedures that have been used to detect fatigue damage. The article suggested that different techniques were suitable for different material systems. For example, the potential drop method worked well with metal fatigue damage, but this technique did not apply to composite materials. All the methodologies were associated with drawbacks such as the acoustic emission technique was prone to an immense amount of noise and the ultrasonic method proved challenging to detect cracks that were parallel to the signal direction. One of the most common limitations of all the techniques lies in the accessibility of the specimen surface. To mitigate the limitation, in-situ crack initiation investigations are reported in several works [24], [25], [26].

Visualization of crack formation is important as it enables researchers to better understand the role of pores and defects in additively manufactured specimens. The availability of scanning electron microscopes [27], transmission electron microscopes [28], [29], and computed tomography facilities [30], [31] has allowed in-situ fatigue investigation of different material systems. These diverse techniques assist in visualizing the potential origins and pathways of crack formation and propagation, respectively. For example, Qian et al. conducted an in-situ investigation inside a scanning electron microscope (SEM). The results showed that anisotropy of defects existed in the vertically fabricated L-PBF Ti-6Al-4V specimens, resulting in lower fatigue strength than horizontally built specimens [25]. Xu et al. reported similar observations for L-PBF AlSi10Mg alloy where fatigue strength was higher in horizontally fabricated specimens [32]. Their in-situ investigation illustrated that the molten pool boundaries were the strain concentration locations and the molten pool varied with the build direction. As a result, the horizontal build direction resulted in a fine microstructure that possessed excellent crack resistance. Furthermore, Wang et al. performed an in-situ fatigue investigation of selective laser-melted Ti-6Al-4V inside an SEM which led to the observation of microcrack nucleation and growth. [33]. The crack initiation and propagation were influenced by the microstructure whereas the persistent slip bands dictated the crack propagation direction. The SEM investigation showed that the grain boundaries hindered the crack growth, and the formation of secondary cracks absorbed a lot of energy which depreciated the growth rate of primary crack. In addition, Xu et al. showed that intergranular fracture was prominent for fatigue short cracks [34]. However, with the increase in crack length, intragranular crack propagation was found to be predominant. From all the works, it can be inferred that SEM observations provided ideas about the role of grain and grain boundaries in fatigue crack growth. However, SEM is not capable of detecting any fatigue crack originating from sub-surface pores or lack-of-fusion pores which is quite common for AM specimens.

Computed tomography (CT) highlights the role of pores and defects in fatigue initiation. Many studies have been conducted to observe the effect of lack-of-fusion pores on fatigue crack initiation and growth [35], [36], [37]. For example, Qian et al. [26] performed an in-situ investigation of fatigue crack growth in L-PBF AlSi10Mg specimens fabricated by selective laser melting technique. The high-cycle fatigue tests were performed inside an X-ray CT facility where the critical lack-of-fusion defect was identified. Higher stress amplitudes produced multiple crack initiations from sub-surface lack-of-fusion defects, and the cracks competed to form the primary crack that led to the specimen's failure. The in-situ investigation demonstrated that during the growth of two off-plane cracks, the overlap of the cracks gave rise to ridges on the fractured surface. Conversely, Wu et al. [38] demonstrated that if sub-surface lack-of-fusion pores were absent, cracks were found to initiate from microstructural features of outer surfaces. Multiple cracks which resulted in crack coalescence influencing the crack propagation path and fracture morphology. Although CT imaging proved to be an effective crack initiation visualization technique for L-PBF specimens, the limitation lies in the scanning resolution of the technique. It is challenging to identify a crack extension below the equipment's imaging resolution. High-resolution CT imaging is always associated with an appreciable amount of time and money, which can be reduced by employing analytical techniques. Furthermore, the CT imaging technique is prone to image quality deterioration by surrounding noise, which can be addressed by analytical methodologies.

Analytical methodologies are reported in the open literature to investigate crack initiation in different materials, but the techniques are mostly applied to traditionally fabricated specimens. For example, Lykins et al. [39] employed several crack initiation parameters such as strain-life parameter, corrected maximum strain, corrected maximum principal strain, Smith-Watson-Topper parameter, critical plane Smith-Watson-Topper parameter, Fatemi-Socie parameter, and Ruiz parameter, to predict the initiation life and crack initiation location in traditionally built Ti-6Al-4V specimens. The results showed that the maximum strain amplitude was the most significant parameter for evaluating initiation life and location. On the other hand, Murthy et al. [40] combined different analytical methodologies to predict the initiation life of cast 304L stainless steel specimens with a notch to induce crack initiation. The work showed that the stress concentration factor proved to be the most reliable predictive method. Among different analytically obtained parameters, the evolution of material stiffness proved to be impactful in monitoring crack initiation in Al7075-T6 as reported in the work of Dharmadhikari et al. [41]. Much of the analytical methodology, in the open literature, was focused on conventionally manufactured components. Moreover, these methodologies included a variety of parameter evaluations along with their calibration for each new material system to identify crack initiation in a component.

The crack initiation investigation becomes more challenging in the case of AM components because AM processes such as the L-PBF technique consist of inherent uncertainties in the process parameters [42] that lead to variation in the microstructure, defects such as lack of fusion, and surface roughness [43]. Several works exist in the open literature, that investigate crack initiation in L-PBF components [44], [45], [46]. For example, Dharmadhikari et al. combined raw experimental data with confocal microscope image information to build a classification model that could predict the initiation life of L-PBF AlSi10Mg specimens [47]. From the experimental data, parameters such as stiffness and energy dissipation were calculated and utilized as inputs to the model. The prediction showed an accuracy of 90% with stiffness as the classification parameter. However, the work was limited to one material, and it lacked an understanding of how the stiffness can pinpoint crack initiation in a specimen. Furthermore, the work was associated with an appreciable amount of data and computational cost. Compliance was used as a parameter to monitor crack initiation in L-PBF specimens as reported in the work of Sheridan et al [48]. Recently, Pal et al. showed that compliance can be used to pinpoint crack initiation in ASTM E606 L-PBF fabricated Hastelloy X specimens [49]. The work involved strain-controlled low cycle fatigue investigation and the results showed that the compliance technique agrees with fractography analysis with an accuracy of 95%. However, the works did not provide a detailed analysis about why and how the material compliance or stiffness evolved with fatigue loading. Furthermore, it is important to demonstrate that the technique is independent of the specimen's geometry, material, fabrication process, and experimental conditions.

To address these research gaps, the current work dives into in-situ low cycle fatigue investigations of miniature specimens. Two different material systems: AlSi10Mg and SS316L, fabricated using L-PBF, are considered in this work to show that the methodology is material-independent. Uniaxial fatigue investigations are conducted in a custom tensile-fatigue testing apparatus with different displacement amplitudes and a positive loading ratio. From the experiments, load-displacement data is collected, and the crack initiation life is evaluated from the data using a compliance-based methodology. Fractography results reveal the crack initiation locations in the specimens and the CT observations corroborate the fractography results. The initiation life evaluation is validated with observations from CT imaging and an analytical load drop calculation. The results show that the maximum difference between compliance-based initiation life evaluation and CT-imaged initiation life observation is 4.8%. Therefore, the proposed methodology can be reliably applied to fatigue investigations for the evaluation of fatigue initiation life.

## 2. Methodology

### 2.1 Specimen fabrication

The specimen geometry is designed to conform to ASTM 25 [50] standard which recommends maintaining the aspect ratio of a specimen greater than 0.2 and the slimness ratio between 4 and 11. Owing to these specifications, the miniature specimen is designed as shown in Figure 1(a). The figure depicts that the gage cross-sectional area is 1 mm² and the gage length is 8 mm. It implies that the aspect ratio is 1 and the slimness ratio is 8. The SolidWorks geometry is used to fabricate the specimens by the L-PBF technique. Two different alloy systems are employed to manufacture the specimens: AlSi10Mg and SS316L.

AlSi10Mg specimens are fabricated using recycled powder from 3D systems. The powder composition is 86.83 wt. % Al, 11.83 wt. % Si, 0.34 wt. % Mg, 0.96 wt. % O, and 0.04 wt. % Fe. The specimens are fabricated using a ProX-320 (3D Systems, Rock Hill, South Carolina, USA) L-PBF equipment at the Center for Innovative Materials Processing through Direct Digital Deposition (CIMP-3D) at Penn State. The parameters used for fabricating the specimens are as follows: layer thickness of 60 $\mu m$, laser power of 325 W, and scanning speed of 1,400 mm/s. A striped hatch design pattern with a hatch spacing of 82 $\mu m$ is used. The specimens used in the experiments are in as-built condition (Figure 1(b)). SS316L specimens (Figure 1(c)) are fabricated using a Concept Laser M2 L-PBF machine located at the Lawrence Livermore National Laboratory. The fabrication process is carried out with a 370 W laser having a beam diameter of 130 $\mu m$, scanning at a speed of 1,350 mm/s. The laser followed a hatch spacing of 90 $\mu m$. SS316L blocks are fabricated, from which the specimens, with a thickness of 1 mm, are extracted by machining.

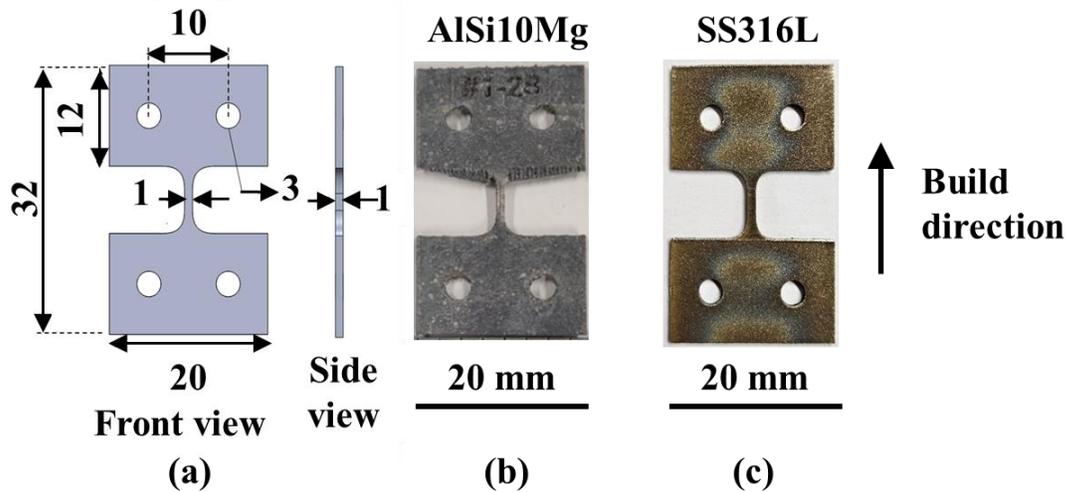

**Figure 1. (a) SolidWorks drawing of the specimen (all dimensions in mm), (b) as-built AlSi10Mg specimen, and (c) SS316L specimen, with their build direction.**

## 2.2 Experimental investigation

The specimen fabrication is followed by conducting fatigue experiments using a custom tensile-fatigue testing apparatus illustrated in Figure 2. In this setup, displacement-controlled tests are carried out utilizing an Ultramotion Servo Cylinder from Honeywell, USA, with a maximum stroke of 25.4 mm. The actuator is connected to the specimen through a grip, where the specimen is secured at the bottom by the lower grip, which also functions as a link to the load cell. The load cell employed in the apparatus is a Model 47 Ultra Precision Fatigue Rated Load Cell from Honeywell, USA, capable of measuring ±1,000 N with a resolution of 0.2 N. Both the actuator and the load cell are connected to a National Instruments Data Acquisition system (DAQ, USB-6251 (BNC) NI-9485 [NI, USA]) having a maximum sampling frequency of 100 kHz. Since the actuator input and load cell output operate in voltages, calibration is necessary. The values are converted to displacement (in mm) for the actuator input and load (in N) for the load cell output, facilitating the assessment of engineering strain and stress. Engineering strain is determined by the ratio of a specific displacement to the gage length (8 mm). A comprehensive explanation of the apparatus's functionality can be found in existing literature [51].

For each material system, five fatigue tests are conducted at different displacement amplitudes but at the same mean displacement. All the fatigue investigations are performed at a frequency of 1 Hz. The AlSi10Mg specimens are subjected to a mean displacement of 0.35 mm and displacement amplitudes of 0.065, 0.070, 0.075, 0.080, and 0.090 mm. On the other hand, SS316L specimens experience a mean displacement of 0.5 mm and displacement amplitudes of 0.075, 0.08, 0.095, 0.105, and 0.115 mm. The specimen identifications along with their respective experimental conditions are described in Table 1. The displacement input curves of A1 and S1 specimens are presented in Figure 3(a), which shows that the mean displacements for AlSi10Mg and SS316L specimens are 0.35 mm and 0.5 mm, respectively. The selection of mean displacement is based on the ultimate tensile strength (UTS) of the materials. A mean displacement of 0.5 mm would have been very close to the UTS of AlSi10Mg, observed in Figure 3(b), whereas a mean displacement of 0.35 mm would have been in the elastic regime for the SS316L specimens. To ensure proper low cycle fatigue investigation, the different mean displacements are selected for different material systems. The raw data from the experiment consists of load, displacement, and cycles to failure. From the load-displacement data, compliance is evaluated, which is further discussed in section 2.3.

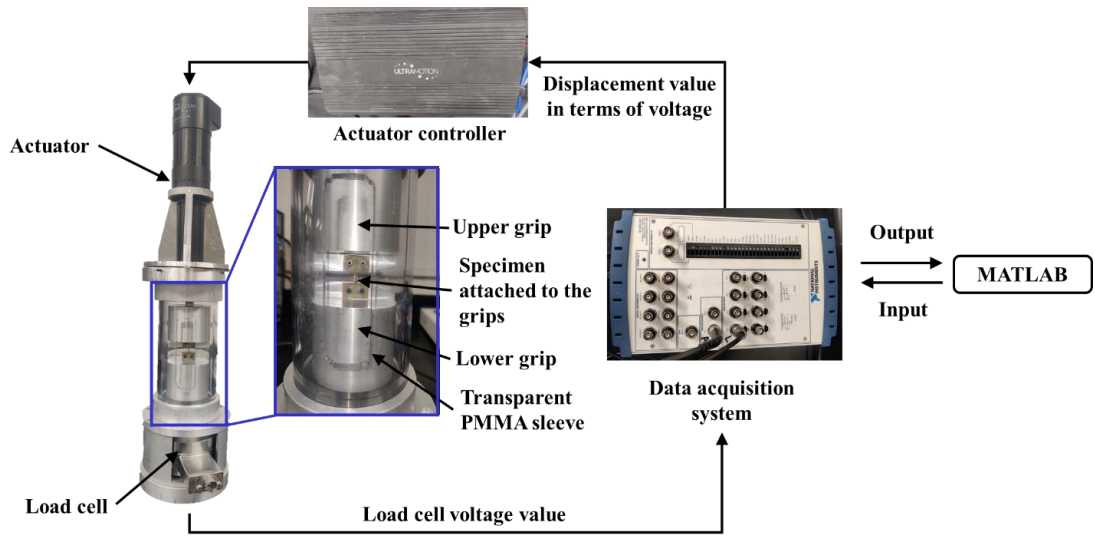

**Figure 2. The fatigue testing apparatus with actuator and load cell connections.**

**Table 1. Experimental conditions of AlSi10Mg and SS316L specimens with specimen identification.**

| AlSi10Mg specimens | | | SS316L specimens | | |
|---|---|---|---|---|---|
| Specimen id | Mean displacement (mm) | Displacement amplitude (mm) | Specimen id | Mean displacement (mm) | Displacement amplitude (mm) |
| A1 | | 0.065 | S1 | | 0.075 |
| A2 | | 0.070 | S2 | | 0.080 |
| A3 | 0.35 | 0.075 | S3 | 0.50 | 0.095 |
| A4 | | 0.080 | S4 | | 0.105 |
| A5 | | 0.090 | S5 | | 0.115 |

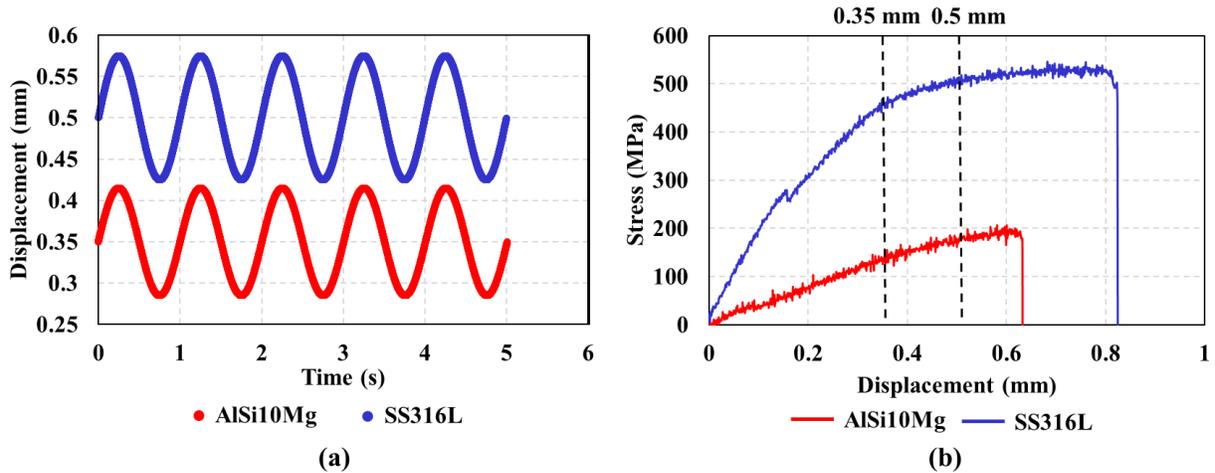

**(a)**                                                            **(b)**

**Figure 3. (a) Sinusoidal displacement inputs for A1 AlSi10Mg specimen and S1 SS316L specimen. (b) Tensile stress-displacement curves for AlSi10Mg and SS316L specimen.**

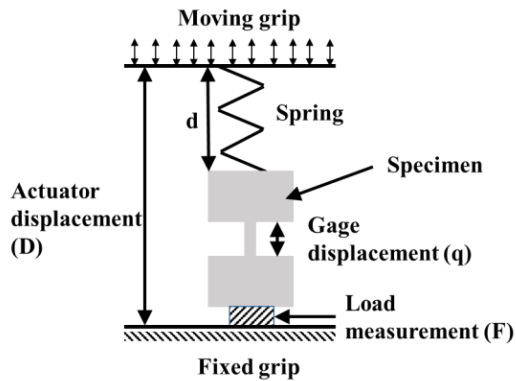

**Figure 4. Schematic of the testing apparatus with all the displacement and loads that the specimen is subjected to during the experiment.**

### 2.3 Compliance analysis

Raw load (F)-displacement (D) data, obtained during fatigue testing, is converted to compliance (c) values as shown in equation (1). For each loading cycle, there is a compliance value associated with the peak load and peak displacement values. Here, peak load is considered because a crack always opens at the peak load of the loading cycle [52]. From compliance-cycles data, the absolute magnitude of the difference in compliance values ($c_2$-$c_1$) is evaluated between two consecutive cycles. These compliance difference values imply the change in compliance that may correspond to crack initiation in the specimen. In this work, the calculation of compliance difference is different from the recent work of Pal et al. [49] because the current work involves displacement-controlled tests, unlike strain-controlled experiments of Pal et al. In displacement-controlled tests, the compliance-displacement slope is not helpful as the peak displacement is always the same. Therefore, the compliance difference between consecutive cycles is evaluated to obtain initiation life. For representation, the compliance difference values are plotted against cycles as discussed in section 3.1.

$$c = D/F \tag{1}$$

In order to validate the compliance technique, an analytical methodology is adopted from the work of Sumpter [53]. Figure 4 is a schematic representation of the loads and displacements that appear during a uniaxial fatigue investigation. The schematic picture consists of a moving grip at the top controlled by an actuator. An actuator displacement, D creates a force or load, F, and a specimen gage displacement, q. The force F is also responsible for other elastic deflections in the apparatus. All these elastic displacements or deformations (d) are lumped in an imaginary spring with stiffness K below the moving grip. Besides the imaginary spring, the specimen has a stiffness of k. As soon as a crack initiates in the specimen, specimen stiffness drops which gives rise to the following situations: a decrease in F, and an increase in d. However, the actuator displacement is fixed due to the displacement-controlled nature of the experiment. The values of the displacements and the force before crack initiation are with suffix 1, and after crack initiation are with suffix 2. Therefore, the following holds: (i) specimen displacements before and after crack initiation are $q_1$, and $q_2$, respectively; (ii) forces before and after crack initiation are $F_1$, and $F_2$, respectively; (iii) spring displacements before and after crack initiation are $d_1$, $d_2$, respectively; and (iv) specimen stiffness before and after crack initiation: $k_1$, $k_2$. All the parameters are tabulated in Table 2. Crack advancement reduces the specimen stiffness ($k_2$<$k_1$) which results in the following equations. The formulation starts with the spring analogy where the force on the specimen can be equated to the product of specimen displacement and specimen stiffness, and it should be equal to the product of imaginary spring stiffness and spring displacement. The equations are based on a fixed actuator displacement owing to displacement-controlled tests.

$$F_1 = k_1 q_1 = K d_1, \tag{2}$$

$$F_2 = k_2 q_2 = K d_2, \tag{3}$$

$$q_2 - q_1 = d_2 - d_1, \; with \; D \; fixed. \tag{4}$$

The equation (4) suggests that the expansion of the spring is equal to the specimen deformation. Equations (2), (3), and (4) can be combined to obtain the following relationships such as,

$$\Delta F = K \Delta q, \tag{5}$$

$$normalized \; load \; drop, \frac{\Delta F}{F_1} = \frac{K \Delta q}{k_1 q_1}, \tag{6}$$

$$\frac{\Delta F}{F_1} = 1 - \frac{k_2}{k_1}\left[\frac{K+k_1}{K+k_2}\right], \tag{7}$$

$$\frac{\Delta F}{F_1} = 1 - \frac{c+c_1}{c+c_2}. \tag{8}$$

Here, $\Delta F = F_1 - F_2$ is the load drop during crack initiation, $\Delta q = q_2 - q_1$ is the increase in gage displacement during crack initiation, $c_1$ and $c_2$ are specimen compliance before and after crack initiation, and C is the apparatus' compliance. The compliance of the apparatus (0.000615 mm/N) is obtained from the work of Pal et al [51]. The apparatus compliance and the specimen compliances are employed in equation (8) to obtain the normalized drop value. To validate the working of the compliance-based methodology, the normalized load drop magnitude from equation (8) must agree with the experimental load drop value.

**Table 2. Definition of the parameters used in equations (1) to (8).**

| Parameters | Definition |
|---|---|
| F | Load |
| D | Actuator displacement |
| K | Imaginary spring stiffness analogous to apparatus stiffness |
| $F_1$, $F_2$ | Load before and after crack initiation |
| $d_1$, $d_2$ | Spring displacement before and after crack initiation |
| $k_1$, $k_2$ | Specimen stiffness before and after crack initiation |
| $q_1$, $q_2$ | Specimen gage displacement before and after crack initiation |
| $\Delta q$ | Change in specimen gage displacement due to crack initiation |
| $c_1$, $c_2$ | Specimen compliance before and after crack initiation |
| C | Apparatus compliance |
| $\Delta F$ | Load drop |

In the current work, crack initiation refers to the formation of 'mechanically small cracks' that correspond to the Stage II tensile cracks as reported in the work of Chopra et al [54]. According to the reported literature, the formation of a Stage II tensile mode crack is preceded by the formation of a microstructurally small crack (less than 10 $\mu m$). These microstructurally small cracks originate because of dislocation slips resulting in Stage I shear mode, which is not investigated in this work. Load drop or change in compliance is visible when the microstructurally small cracks transition to mechanically small cracks (ranging from 10 $\mu m$ to 3 mm) or in other words, the cracks transition to Stage II tensile mode from Stage I shear mode. Stage II tensile mode mechanically small cracks are more important than microstructurally small cracks which may not always enter stage II mode as they may be arrested due to some dislocation, but once they transition to Stage II mode, the mechanically small cracks lead to the failure of a specimen.

### 2.4 In-situ investigation in CT

The fatigue investigations are performed inside an X-ray CT facility. Figure 5 shows the fatigue testing apparatus inside the CT facility. The transparent Polymethyl Methacrylate (PMMA) sleeve of the apparatus allows the penetration of the X-rays to image the specimen. CT imaging is carried out in the GE v|tome|x system located at Penn State. The equipment is capable of imaging objects at sub-500 nm resolution. In this work, the specimens are imaged at a voxel resolution of 10 $\mu m$ which is feasible with 200 kV and 75 $\mu A$ of scanning voltage and current, respectively. For each specimen, three projections are recorded at each acquisition angle using the step-scan technique. Within the range of 360°, projections are acquired at an interval of 0.3°. The projection images are reconstructed utilizing the filtered back-projection algorithm and the reconstruction produces a volume ('.vol') file. The volume file is imported into ImageJ [55] with dimensions length, width, and number of cross-sectional images. The images are exported as stacked files in Tagged Image File Format ('.tiff'), which are imported into Avizo 2021 3D software [56] to reconstruct the 3D specimen geometry.

A major step in in-situ fatigue investigation is to create an interval in the experiment that allows the X-ray to image the specimen at a particular instant of time. In this work, the experiments are paused at specific intervals for a time length of 60 minutes which is a requirement of the X-ray CT equipment to achieve the desired resolution of 10 $\mu m$. The imaging is performed until the specimen fails and the scanned images are reconstructed to obtain 3D geometries.

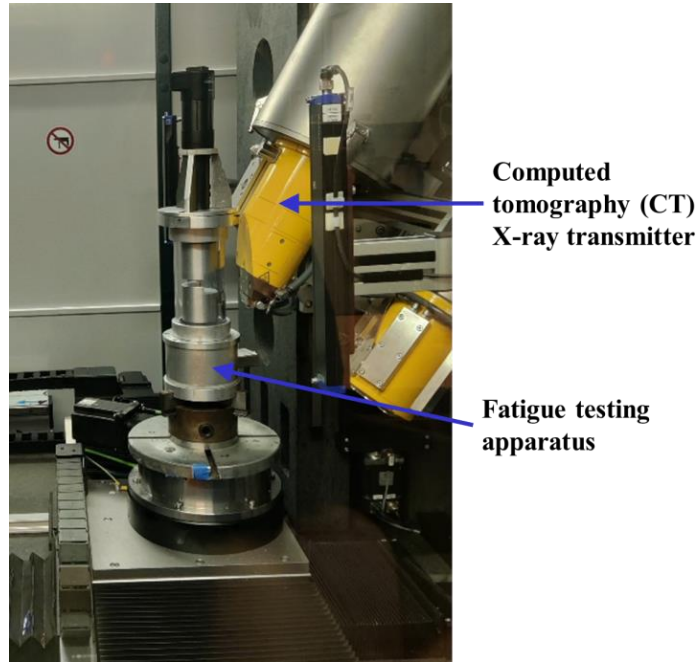

**Figure 5. Representation of the testing apparatus inside the CT facility.**

### 2.4 Fractography analysis

After the fatigue investigations, fractured surfaces of the specimens are investigated by fractography analysis using a Q250 analytical scanning electron microscope (Thermo Scientific, USA). The system provides a range of operating voltages: 200 V to 30 kV, enabling sub-nanometer resolution. Topographic contrast on the fractured surfaces is detected by a secondary electron detector. For both material systems, the fractured surfaces are examined at an operating voltage of 20 kV and a current of 17 nA. These operating conditions achieve a resolution of 1.3 $\mu m$/pixel.

### 3. Results and discussion

The section discusses the results from in-situ fatigue investigations of AlSi10Mg and SS316L specimens. The results include visualization of crack initiation via CT, analysis of load-displacement data to obtain compliance, and correlation of compliance with CT visualization to obtain initiation life. Additionally, fractography analysis corroborates the analytical and in-situ crack initiation observations. In these sections, the representative specimens for qualitative evaluation are randomly selected and the results of all the specimens are used in quantitative evaluations. Finally, Coffin-Manson-Basquin parameters are extracted from the experimental observations, and they are utilized to correlate initiation life with strain amplitude.

### 3.1 Visualization of fatigue crack evolution in CT

The in-situ observations of a representative AlSi10Mg specimen A2 (displacement amplitude: 0.070 mm) are presented in Figures 6(a) and 6(b). Figure 6(a) shows the CT-imaged gage area of the specimen before and after the fatigue failure. It also depicts the cross-sectional planes used to present the crack initiation in the next set of images (Figure 6(b)). From Figure 6(b), it is observed that the crack initiates from the surface flaw, denoted by 'crack' in the figure. The initiation occurs between the $5,000^{th}$ and $5,500^{th}$ cycles for this specimen. The CT imaging of the AlSi10Mg specimen demonstrates that the crack propagation occurs for a small duration leading to a very rapid failure after crack initiation. This observation is associated with the brittle nature of AlSi10Mg specimens, which do not undergo significant plastic deformation before the final fracture [57]. In addition to the material characteristics, the presence of a considerable number of inclusions and porosities in the specimen reduces the load-bearing capacity [58], [59]. As a result, after the crack initiation, the crack reaches a point where the specimen is overstressed, and rapid fatigue failure occurs across the specimen [60]. Another interesting fact, from Figure 6(b), is that the rapid failure path includes the porosities near the crack initiation. Furthermore, the CT images reveal the presence of a substantial number of pores in the specimen, and the circular shape of the pores implies that the pores arise from trapped gases during the fabrication [61].

A similar crack initiation phenomenon is observed for the A1 specimen which is investigated at a displacement amplitude of 0.065 mm. Figure 6(c) depicts the region of interest in the gage section at two distinct stages of fatigue investigation. The longitudinal cross-section images of the 3D geometry in Figure 6(d) reveal the crack initiation from a surface flaw at the $7,750^{th}$ cycle. The crack initiation is understood by comparing the images from the four CT imaging results in Figure 7(d). The first two images correspond to the $0^{th}$ cycle and $7,500^{th}$ cycle, and they have the same length of the surface flaw, marked in green. However, the crack extends from the flaw between $7,500^{th}$ and $7,750^{th}$ cycles, implying that the crack initiates in that period. Apart from the surface flaw, a considerable number of gas pores can be observed in the specimen, similar to the A2 specimen. The specimen failed around 8,500 cycles with a propagation life of around 800 cycles, owing to the brittle nature of AlSi10Mg specimens. For both AlSi10Mg specimens, the crack extensions from the existing surface flaws are measured from the CT images. The measurements show a crack extension of 0.05 mm between the $4,750^{th}$ and $5,000^{th}$ cycles for the A2 specimen, and 0.02 mm between the $7,500^{th}$ and $7,750^{th}$ cycles for the A1 specimen. The extension measurements are limited by the CT imaging resolution of 10 $\mu m$.

The in-situ fatigue investigations of SS316L specimens show significant propagation until final failure. The observation of specimen S3 is presented in Figures 7(a) and 7(b) where the initiation occurs around 7,000 cycles, and the propagation continues up to 9,000 cycles. Figure 7(a) demonstrates the CT imaged gage area and Figure 7(b) indicates the presence of major lack-of-fusion pores that can act as crack initiation sites during the fatigue loading. The lack-of-fusion pores are identified by their non-circular shape and their location in the specimen [62]. The trailing edge of the pores is perpendicular to the build direction which implies that the pores are a result of a lack of fusion between two consecutive layers during the fabrication process. The lack-of-fusion pores are close to the surface, and with the progress of the experiment, the crack is found to initiate from the top lack-of-fusion pores, leading to the final failure of the specimen. The crack initiation refers to the extension of the lack-of-fusion pore by 0.02 mm. Besides lack-of-fusion pores, there are circular gas pores inside the specimen that do not take part in fatigue crack initiation.

The CT imaging results of another SS316L specimen are presented in Figures 7(c) and 7(d). The images present the fatigue investigation of the S5 specimen at a displacement amplitude of 0.115 mm. Figure 7(c) shows the difference between 3D reconstructed geometries of the gage region of the specimen at the $0^{th}$ cycle and the $5,500^{th}$ cycle. The lack-of-fusion pore in the middle of the specimen enlarges during the fatigue loading, primarily leading to specimen failure. The crack extension from the lack-of-fusion pore is found to be 0.03 mm at the $5,500^{th}$ cycle. The longitudinal cross-section images of the 3D geometry in Figure 7(d) aid in the visualization of the crack initiating from the lack-of-fusion pore and propagating through the interior of the specimen. The lack-of-fusion pores extend between the 3,500ᵗʰ and 5,500ᵗʰ cycles, which suggests the crack has initiated in that period. The initiation life of the specimen is considered 5,500 cycles from CT imaging as the crack initiation is more visible in that cycle from the cross-sectional images. There are other lack-of-fusion pores in the specimen, but primary crack growth is favored by the lack-of-fusion pores marked in green.

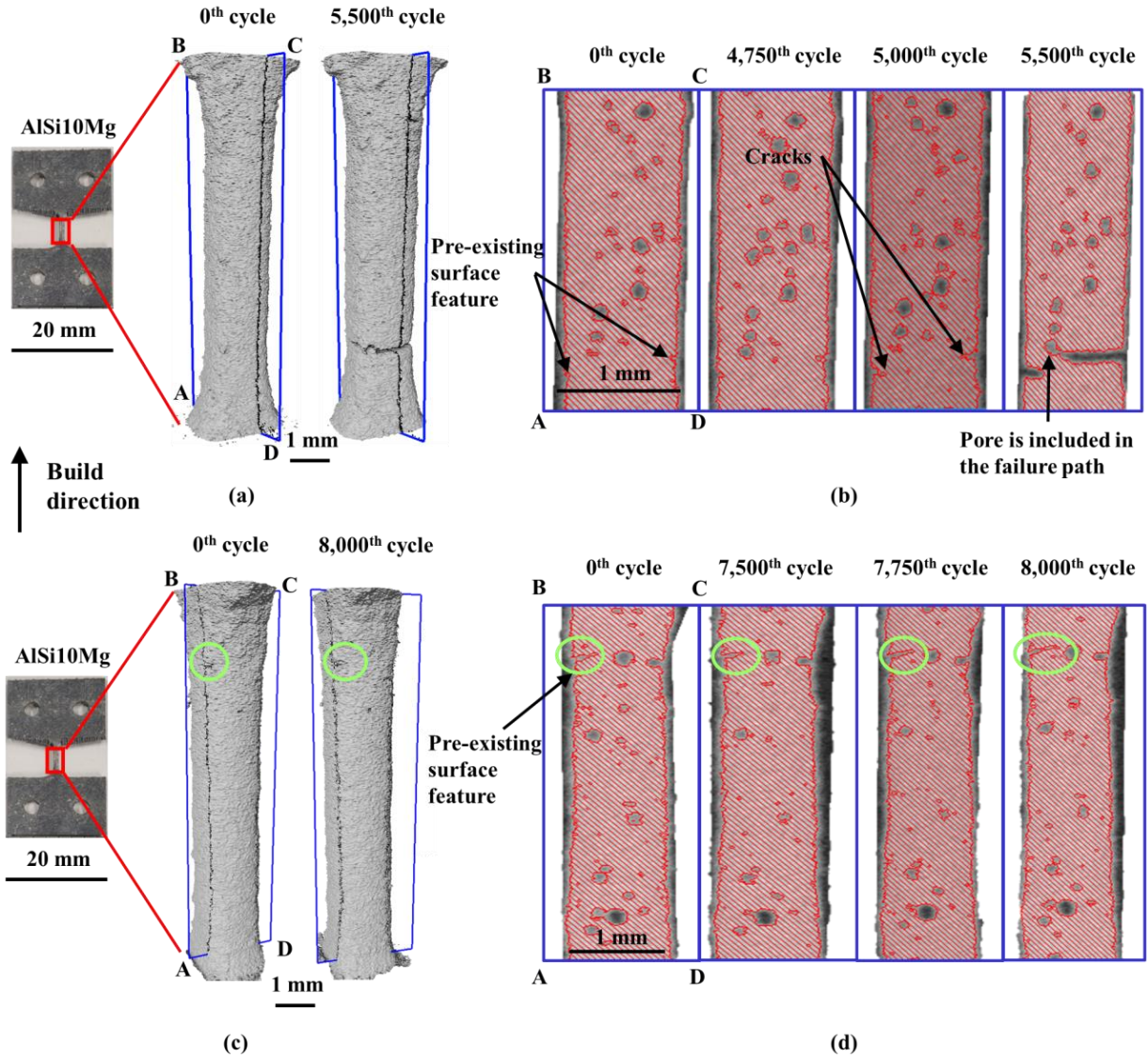

**Figure 6. (a) CT-imaged portion of representative AlSi10Mg specimen A2 at 0ᵗʰ and 5,500ᵗʰ cycles. The blue rectangle (ABCD) indicates the longitudinal cross-section plane used to demonstrate in-situ crack initiation and propagation in the next set of images. (b) In-situ crack growth images of AlSi10Mg specimen at various stages of fatigue life. Sub-surface pores are present in the specimen from which crack starts to grow at around 5,000ᵗʰ cycle and the specimen fails at a fatigue life of 5,500 cycles. (c) The CT-imaged portion of the A1 AlSi10Mg specimen at 0ᵗʰ and 8,000ᵗʰ cycles. The blue rectangle (ABCD) indicates the longitudinal cross-section plane used to demonstrate in-situ crack initiation in the next set of images. (d) In-situ crack growth images of A1 AlSi10Mg specimen at various stages of fatigue life. Surface flaws (marked by a green ellipse) are present in the specimen**

**from which the crack starts to grow at around the 7,750ᵗʰ cycle. The red-hatched region shows the interior cross-section of the specimen that is separated from pores and cracks.**

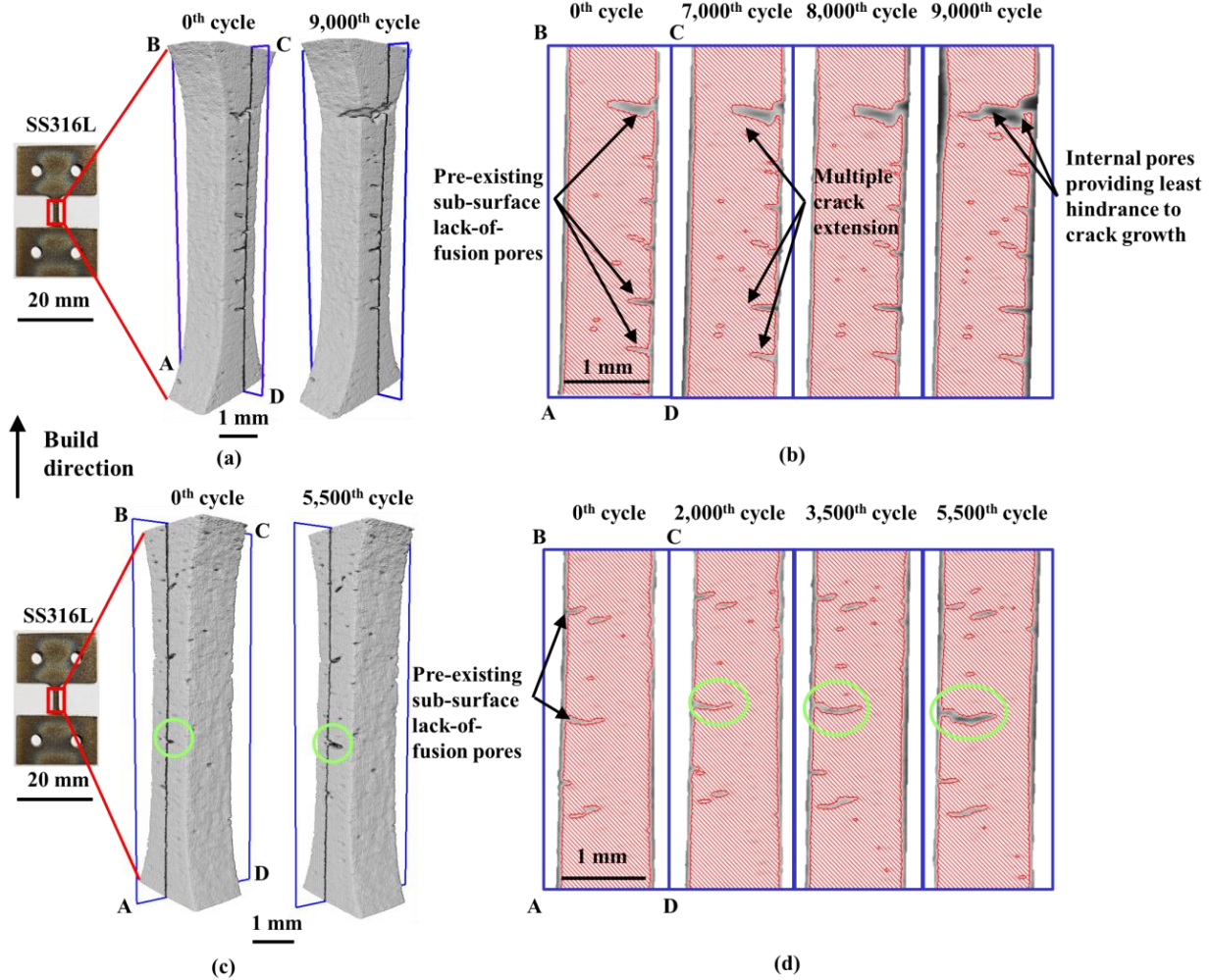

**Figure 7. (a) CT-imaged portion of a representative SS316L specimen S3 at 0ᵗʰ and 9,000ᵗʰ cycles. The blue rectangle (ABCD) indicates the longitudinal cross-section plane used to demonstrate in-situ crack initiation and propagation in the next set of images. (b) In-situ crack growth images of SS316L specimen at various stages of fatigue life. Sub-surface pores are present in the specimen from which crack starts to grow at around 8,000ᵗʰ cycle and the specimen fails at a fatigue life of 9,000 cycles. (c) CT-imaged portion of S5 SS316L specimen at 0ᵗʰ and 5,500ᵗʰ cycles. The blue rectangle (ABCD) indicates the longitudinal cross-section plane used to demonstrate in-situ crack initiation in the next set of images. (d) In-situ crack growth images of S5 SS316L specimen at various stages of fatigue life. The crack is observed to be extended from a lack-of-fusion pore (marked by a green ellipse) at the 5,500ᵗʰ cycle. The red-hatched region shows the interior cross-section of the specimen that is separated from pores and cracks.**

From the CT imaging of SS316L specimens, a significant amount of plastic deformation is observed which leads to a noticeable crack propagation period [63], [64]. The high gradient of plastic deformation causes the initial crack tip (Figure 7(b)) to become a highly distorted, forming a rounded crack tip known as a 'blunt' crack [65]. During the formation of the blunt crack, a minimal amount of crack growth takes place.

After the blunt crack formation, high stress levels for the reduced area, along with the notch effect of the crack, result in stable crack growth [66], [67]. A stable crack growth refers to crack extension with each loading cycle. The crack extends in the horizontal direction with the subsequent loading cycles, abiding by the crack propagation path for Mode I loadings [68]. Due to the presence of internal pores along the crack path (Figure 7(b)), the crack propagation direction changes by following the path of least hindrance. The crack is then at a different angle with respect to the loading direction which causes the crack to follow the path of mixed-mode loading [69]. The crack propagation path encounters another internal pore that further changes the crack propagation angle, leading to the failure of the S3 specimen. At the 9,000th cycle, it can be observed that an appreciable amount of specimen distortion has taken place due to crack propagation. The crack extensions from the lack-of-fusion pores are measured from the CT images. The measurements suggest that the crack extends by 0.02 mm around 7,000th cycles for the S3 specimen, and 0.03 mm between the 3,500th and 5,500th cycles for the S5 specimen. One significant difference between AlSi10Mg and SS316L specimens lies in the surface roughness of AlSi10Mg specimens. Therefore, from all the CT images, an interesting observation is that crack initiation in each AlSi10Mg specimen takes place from surface flaw, whereas in SS316L specimens, crack onset occurs from a sub-surface lack-of-fusion pore. As the internal pores do not play a role in the crack initiation in a specimen, statistics of internal pores in the specimens are not reported in this article.

## 3.2 Evolution of compliance during fatigue testing

The fatigue investigation provides load-displacement data which is converted to compliance. Compliance difference between each cycle is evaluated and the variation of compliance difference is plotted with the consecutive cycles to locate the crack initiation cycle during the experiment. In this methodology, a cycle is considered the crack initiation cycle when the compliance difference values exceed 0.00004. This threshold is chosen because values lower than 0.00004 can be attributed to instrument noise, typically resulting from load variations of 4 to 5%. This definition is further validated by fractography and CT imaging in sections 3.3 and 3.4. Figures 8(a) and (b) show the load-cycle curve and the compliance difference-consecutive cycles curve of the A2 AlSi10Mg specimen, respectively. As the compliance difference is calculated between two consecutive cycles, significantly large values are observed for the 5,278th-5,279th cycles and 5,279th-5,280th cycles, shown in the inset picture of Figure 8(b). This implies that the 5,279th cycle is the probable crack initiation cycle because the load drop at that cycle increases the compliance value. Therefore, the compliance difference values for the 5,278th-5,279th cycles and the 5,279th-5,280th cycles are larger than the other consecutive cycles. At the time of crack initiation, a load drop occurs across the direction transverse to crack initiation [70] which is the loading direction in this case. From the load-cycle curve, the load drop is not significantly visible to mark the crack initiation instance. However, it becomes prominent with the evaluation of compliance differences as it incorporates load variation between the time of crack initiation and after crack initiation. For further validation, compliance values at 5,278 and 5,279 cycles are substituted in equation (8), and the normalized load drop value is found to be 0.066 by the following calculation,

$$\frac{\Delta F}{F_1} = 1 - \frac{C+c_1}{C+c_2} = 1 - \frac{C+c_{5278}}{C+c_{5279}} = 1 - \frac{0.000615+0.0034}{0.000615+0.0035} = 0.066. \tag{9}$$

This value matches the experimental normalized load drop value of 0.07 which is evaluated by,

$$\frac{\Delta F}{F_1} = \frac{F_{5278}-F_{5279}}{F_{5278}} = \frac{126-117}{126} = 0.07, \tag{10}$$

indicating that the 5278th cycle is the crack initiation cycle. Furthermore, the load-cycle and the compliance difference-consecutive cycle curves are plotted for the A1 specimen, shown in Figures 8(c) and 8(d).

Similarly, a load drop is not evident from the load-cycle curve, but the compliance difference values show a noticeable change between the 7,456th-7,457th cycles and 7,457th-7,458th cycles, as depicted in the inset picture of Figure 8(d). This suggests that a load drop occurs at the 7,457th cycle which increases the compliance difference values for the 7,456th-7,457th cycles, and 7,457th-7,458th cycles, indicating crack initiation at the 7,457th cycle. Moreover, compliance values at the 7,456th and 7,457th cycles are used in equation (8) to evaluate the normalized load drop, which is around 0.07, calculated by,

$$\frac{\Delta F}{F_1} = 1 - \frac{C + c_1}{C + c_2} = 1 - \frac{C + c_{7456}}{C + c_{7457}} = 0.07. \tag{11}$$

This value matches the experimental normalized load drop value of 0.08 which is evaluated by,

$$\frac{\Delta F}{F_1} = \frac{F_{7456} - F_{7457}}{F_{7456}} = 0.08. \tag{12}$$

This agreement between analytical calculation and experimental observation provides a solid foundation for the compliance methodology.

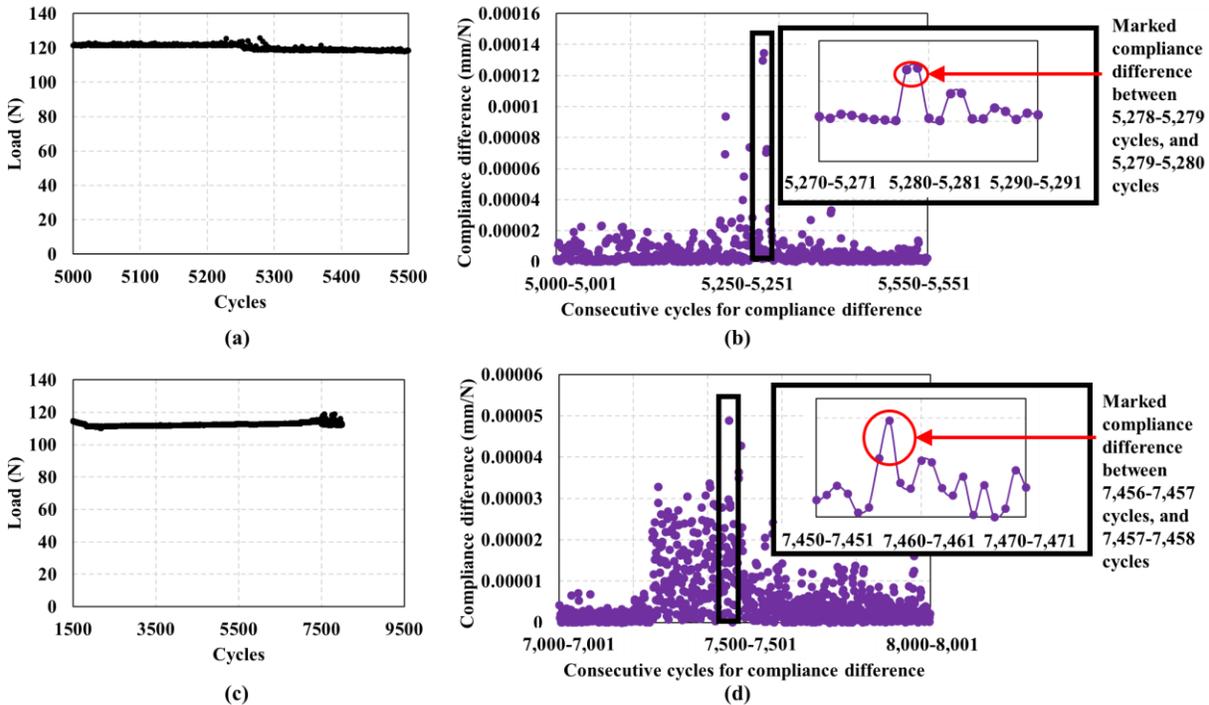

**Figure 8. (a) Load versus cycles plot of AlSi10Mg specimen A2 showing the variation in load with the increase in fatigue cycles, and (b) compliance versus consecutive cycles plot with an inset picture of the area marked by a green rectangle, indicating a marked increase in compliance difference values for specific consecutive cycles. (c) Load versus cycles plot of A1 AlSi10Mg specimen showing the variation in load with the increase in fatigue cycles, and (d) compliance versus consecutive cycles plot with an inset image of the zone marked by a green rectangle, indicating a marked increase in compliance difference values for specific consecutive cycles.**

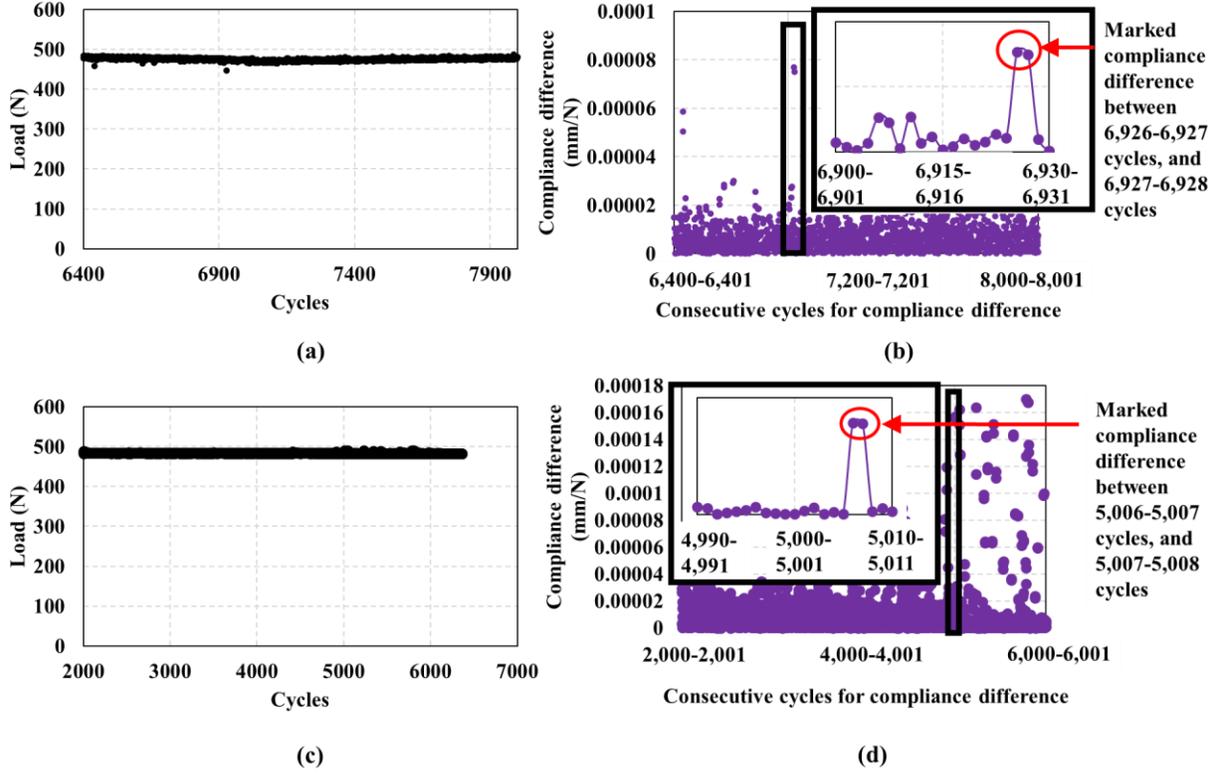

**Figure 9. (a) Load versus cycles plot for S3 SS316L specimen, and (b) compliance versus consecutive cycles plot with an inset picture of the area marked by a green rectangle, indicating a marked increase in compliance difference values for specific consecutive cycles. (c) Load versus cycles plot for S5 SS316L specimen, and (d) compliance versus consecutive cycles plot with an inset picture of the area marked by a green rectangle, indicating a marked increase in compliance difference values for specific consecutive cycles.**

A similar compliance difference evaluation is carried out for SS316L specimens. For SS316L specimens, compliance differences less than 0.00006 are considered instrument noises (load variation of 4 to 5%). Figures 9(a) and (b) show the load-cycle variation and the variation of compliance difference with the consecutive cycles for the S3 specimen, respectively. The compliance difference values rise significantly at 6,926-6,927 cycles and 6,927-6,928 cycles, shown in the inset image of Figure 9(b), indicating a load drop and a probable crack initiation at the 6,927th cycle. To validate the methodology, the normalized load drop is evaluated from the experimental observations to confirm with equation (8). For the S3 specimen, the normalized load drop between the 6,926th cycle and the 6,927th cycle is found to be 0.06, calculated by

$$\frac{\Delta F}{F_1} = \frac{F_{6926} - F_{6927}}{F_{6926}} = \frac{476 - 446}{476} = 0.06, \tag{13}$$

and equation (8) provides a value of 0.04, evaluated by

$$\frac{\Delta F}{F_1} = 1 - \frac{C + c_1}{C + c_2} = 1 - \frac{C + c_{6926}}{C + c_{6927}} = 1 - \frac{0.000615 + 0.001248}{0.000615 + 0.001325} = 0.04, \tag{14}$$

indicating that the 6,927th cycle is the crack initiation cycle. Crack initiation implies the creation of two new surfaces by using the accumulated strain energy in the specimen [71]. For the SS316L specimens, the crack propagation period is more prominent than in AlSi10Mg specimens because SS316L is more ductile than AlSi10Mg. Ductility leads to a significant amount of plastic deformation around the crack tip, causing a change in stress levels during crack propagation [63], [64]. Crack initiation is immediately followed by a gradual decrease in compliance difference values for a brief period known as the 'blunting' stage [65]. At least three different mechanisms are responsible for the change in compliance differences: redistribution of elastic and plastic zones, nucleation and growth of voids, and cross-sectional area reduction due to crack growth [65]. Following the blunting stage, the crack enters the stable growth region where the compliance values remain unchanged. The load-cycle and compliance difference values of the S5 specimen are shown in Figures 9(c) and 9(d), respectively. An increase in compliance difference is observed for the 5,006-5,007 cycles and 5,007-5,008 cycles, presented in the inset picture of Figure 9(d). This implies that the 5,007th cycle might be considered the crack initiation instance. To further validate this observation, the normalized load drop is calculated from equation (8), using the experimental compliance values at the 5,006th and 5,007th cycle. The evaluation results in an analytical normalized load drop value of 0.065, calculated by equation (15), which agrees with the experimental value of 0.067 using equation (16).

$$\frac{\Delta F}{F_1} = 1 - \frac{C+c_1}{C+c_2} = 1 - \frac{C+c_{5006}}{C+c_{5007}} = 0.065, \tag{15}$$

$$\frac{\Delta F}{F_1} = \frac{F_{5006}-F_{5007}}{F_{5006}} = 0.067. \tag{16}$$

Therefore, for both material systems, the compliance methodology results in good agreement with the analytical load drop calculation. All the load drop values are greater than the load cell resolution of 0.2 N.

### 3.3 Correlating compliance with CT observation

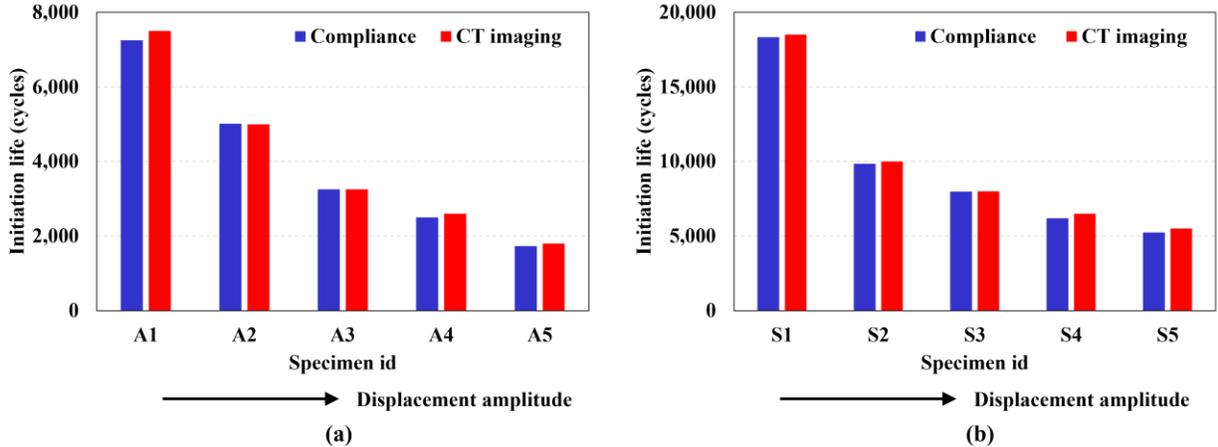

**Figure 10. Comparison between initiation life obtained from compliance technique and CT observations for all (a) AlSi10Mg and (b) SS316L specimens. The specimens are arranged in an increasing order of displacement amplitude. The results from both techniques are in close agreement with each other.**

The initiation life of AlSi10Mg specimens obtained from CT observations is plotted alongside the potential initiation cycle obtained from compliance calculation in Figure 10(a). The bar chart shows that the results from both methodologies are close to each other, with a maximum difference of 4.8%. The primary reason for the difference is that CT imaging is performed at specific intervals, which may capture the exact moment of crack initiation. Therefore, the initiation life from CT observation is generally greater than the value calculated using the compliance technique. Figure 10(b) depicts the results for all SS316L specimens. The maximum difference between the two methods is around 4.7% for SS316L specimens, attributable to the time intervals at which CT imaging is performed. Due to these intervals, CT imaging may detect crack extension or crack growth after the actual crack initiation in the specimen. Moreover, CT imaging is limited by resolution, which may sometimes fail to detect crack initiation at small scales. Hence, the compliance-based methodology is a better technique for pinpointing fatigue crack initiation in a specimen.

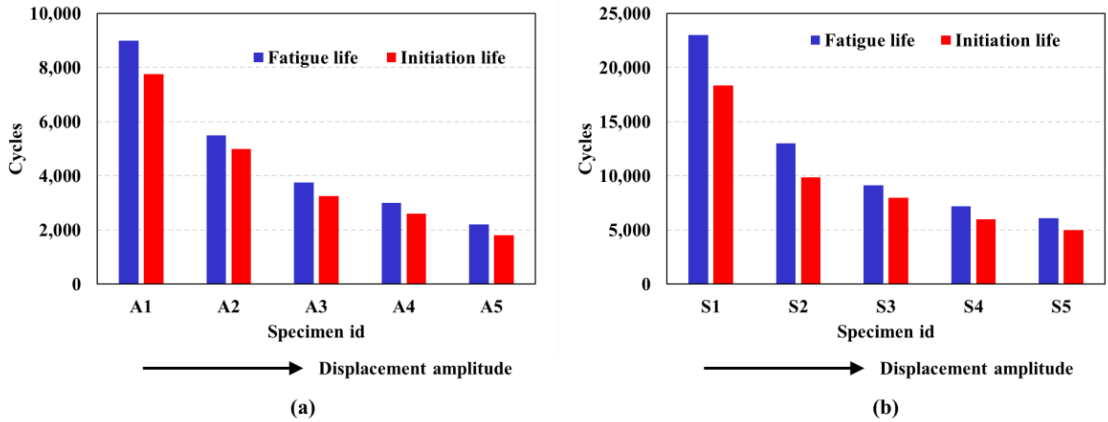

**Figure 11. Fatigue and initiation life variation of (a) AlSi10Mg specimens and (b) SS316L specimens. The specimens are arranged in an increasing order of displacement amplitude. The fatigue life and the initiation life decrease with the increase in displacement amplitude.**

The fatigue life and the initiation life obtained from compliance methodology for all AlSi10Mg specimens are presented in Figure 11(a). The bar chart illustrates that the fatigue lives of the specimens decrease with the increase in displacement amplitude. A higher displacement amplitude results in a higher stress, leading to a lower fatigue life. Furthermore, initiation life is proportional to fatigue life because higher displacement amplitudes induce higher strain accumulation, causing cracks to initiate earlier than in specimens subjected to smaller displacement amplitude values. Since it is well established in the literature that AM specimens show significant scatter in fatigue life [72], two more experiments are conducted with AlSi10Mg specimens under the same conditions as the A3 specimen (displacement amplitude of 0.075 mm). The fatigue life evaluation of those specimens suggests that a variance of 500 cycles is present at that amplitude, implying a difference of 13%.

A similar plot is prepared for SS316L, shown in Figure 11(b), where the initiation life of the specimens is proportional to the fatigue life. The difference between the fatigue and initiation lives is larger for SS316L specimens than for AlSi10Mg specimens. The significant difference can be attributed to the substantial plastic deformation in SS316L before the final fracture as discussed in sections 3.1 and 3.2. Large plastic deformation implies high energy expenditure in the formation of new surfaces during crack propagation [5]. Therefore, a significant amount of crack propagation is associated with SS316L specimens, unlike AlSi10Mg specimens. Like AlSi10Mg specimens, repeatability is investigated for SS316L specimens by performing two additional experiments under the conditions of the S3 specimen (displacement amplitude of 0.095 mm). The fatigue life evaluation of these specimens indicates a variance of 1,000 cycles at that displacement amplitude, corresponding to a percentage difference of 12.5. Thus, these percentage differences of 13 and 12.5 for AlSi10Mg and SS316L specimens, respectively, are typical for low cycle fatigue investigations [73], especially AM specimens [72], [74]. The selection of A3 and S3 specimens for repeatability investigations lies in the displacement amplitudes, which are mid-way between the lowest and highest values for each specimen set.

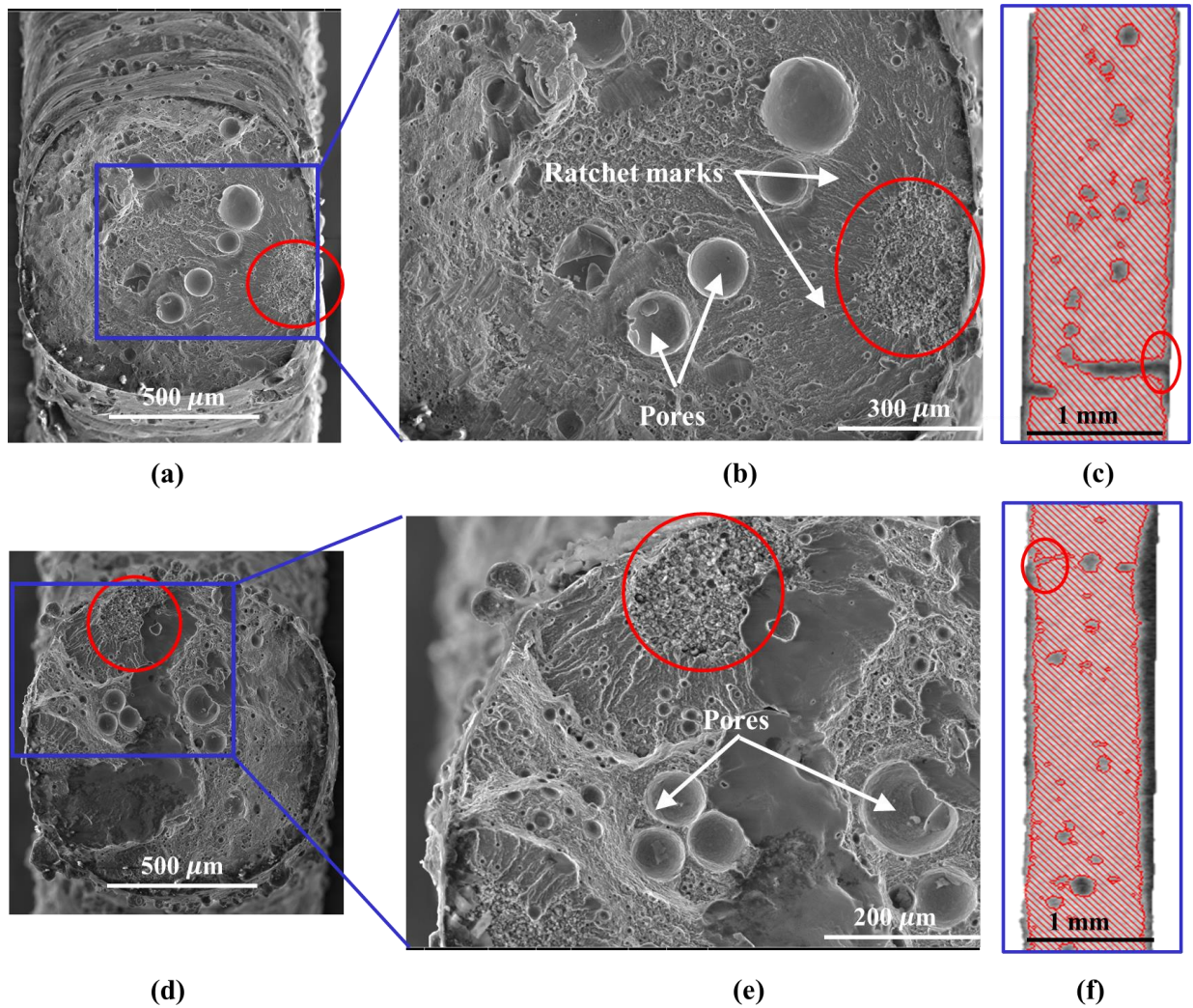

**Figure 12.** Representative fractography image of the A2 AlSi10Mg specimen: (a) the crack initiation zone and (b) a zoomed-in image with ratchet marks and pores. (c) The crack initiation site is correlated with the cross-sectional CT image. Representative fractography image of A1 AlSi10Mg specimen: (d) the crack initiation zone and (e) zoomed-in image with pores. (f) The crack initiation site is corroborated by the cross-sectional CT image. The crack initiation zones are demarcated by the red ellipse on the fractured surfaces.

### 3.4 Fractography analysis

Fractography analysis is performed after the fatigue experiments. The fractography images of A2 and A1 AlSi10Mg specimens in Figure 12 depict visible crack initiation and rapid failure zones with a negligible propagation zone confirming that engineering of process parameters has successfully made the material brittle. In Figures 12(a) and (b), the convergence of ratchet marks in the A2 specimen identifies the initiation zone on the fractured surface. These ratchet marks suggest that the initiation is from high-stress concentration regions or in other words, surface flaws [60]. The fractography image is consistent with the findings of Pan et al. [75], where crack initiation was observed from surface flaws. It is already established in the open literature that surface flaws play a pivotal role in fatigue failure of as-built AlSi10Mg specimens [76], [77]. The initiation zone in the fractography analysis also suggests a possible interaction between surface flaws and inhomogeneous microstructure, which can be further investigated in the future. Moreover, the ratchet marks or ridges arise radially from the initiation zone, depicting crack growth. These ratchet marks around the pores imply that the crack growth is enhanced by those pores. The crack initiation from a surface flaw is corroborated by the observation from CT imaging, shown in Figure 12(c), where the initiation zone is marked by a red ellipse. A similar observation is noted for the A1 specimen, shown in Figures 12(d), (e), and (f). The crack initiation zone (marked by a red ellipse) is identified on the fractured surfaces by the ratchet marks, depicted in Figure 12(e). The ratchet marks converge to a surface flaw, which is the crack initiation site, confirmed by the CT image in Figure 12(f). After the crack initiates from surface flaws, the crack grows rapidly in the neighborhood of pores with a negligible number of striation marks. It is reported in the open literature that vertically as-built AlSi10Mg specimens offer the least resistance to crack growth [24]. This is because the crack grows along the neighborhood of melt pool boundaries, where heat-affected zones provide a less ductile network. The cracks tend to propagate along the weakest links, which are the interlayer heat-affected zones and pores. Moreover, the Si particles along the melt pool boundaries and the surface flaws enhance crack initiation [78]. The absence of dimples from the fractography images reveals the brittleness of the material which implies negligible plastic deformation in the specimen [79]. The negligible crack propagation on fractured surfaces is validated by the small difference of around 300 and 800 cycles between cycles to failure and cycles to crack initiation for A2 and A1, respectively. The brittle behavior of the L-PBF AlSi10Mg specimens is reported in the existing literature [77].

On the other hand, SS316L fractography images show the presence of initiation and propagation zones, confirming that the process parameter engineering is successful in making the material ductile. Figures 13(a), and (b) present the fractography images of the S3 specimen. The propagation zones are identified by the striation marks on the surface as shown in Figure 13(b), and the initiation zone is identified by the plain-faceted region. The initiation occurs from a sub-surface lack-of-fusion porosity oriented perpendicular to the build direction as observed in Figure 13(c). A similar observation was reported by Shreshtha et al [80]. The crack initiates and propagates through the specimen in a perpendicular direction to the cyclic loading. On the fractured surface (Figure 13(b)), stepped crack propagation is observed, corroborated by the CT cross-sectional image in Figure 13(c). The appearance of the stepped crack is attributed to the existence of internal pores along the crack propagation path, altering the angle of crack growth, as depicted in the figure. The striation marks appear due to the changes in the stress levels with each cycle as the crack grows across the specimen [60]. During a tensile loading cycle, expansion of the crack tip leads to the crack growth by a finite amount followed by the crack closure during the compression cycle. The opening, expanding, and closing behavior of the crack leads to the formation of beach marks. The crack initiation and the crack growth behavior in SS316L are further investigated in the S5 specimen. The crack initiates from the sub-surface lack-of-fusion pore as marked in Figure 13(c), (d), and (e). After the crack onset, the crack propagates, and the crack growth leads to a valley-type formation on the fractured surface. Furthermore, due to the ductile behavior of the material, high stress levels induce plastic deformation with accumulated plastic strain energy in the specimen [81]. The accumulated energy is used in the creation of a new surface while the crack propagates. As the crack grows, the stress increases due to the decrease in the load-bearing area in front of the propagated crack. When the area is overstressed, rapid failure occurs across the specimen. The striation marks suggest a significant crack propagation period which is further corroborated by the difference of 1,000 and 1,500 cycles between cycles to failure and cycles to initiation for S3 and S5 specimens, respectively.

Both material systems are subjected to a similar range of strain levels in the plastic regime. However, the crack propagation period is significantly different between the two material systems. AlSi10Mg specimens show a propagation period ranging from 300 to 800 cycles, whereas SS316L specimens demonstrate a propagation period ranging from 1000 to 1500 cycles. This can only be attributed to the distinct material behavior during fatigue investigations.

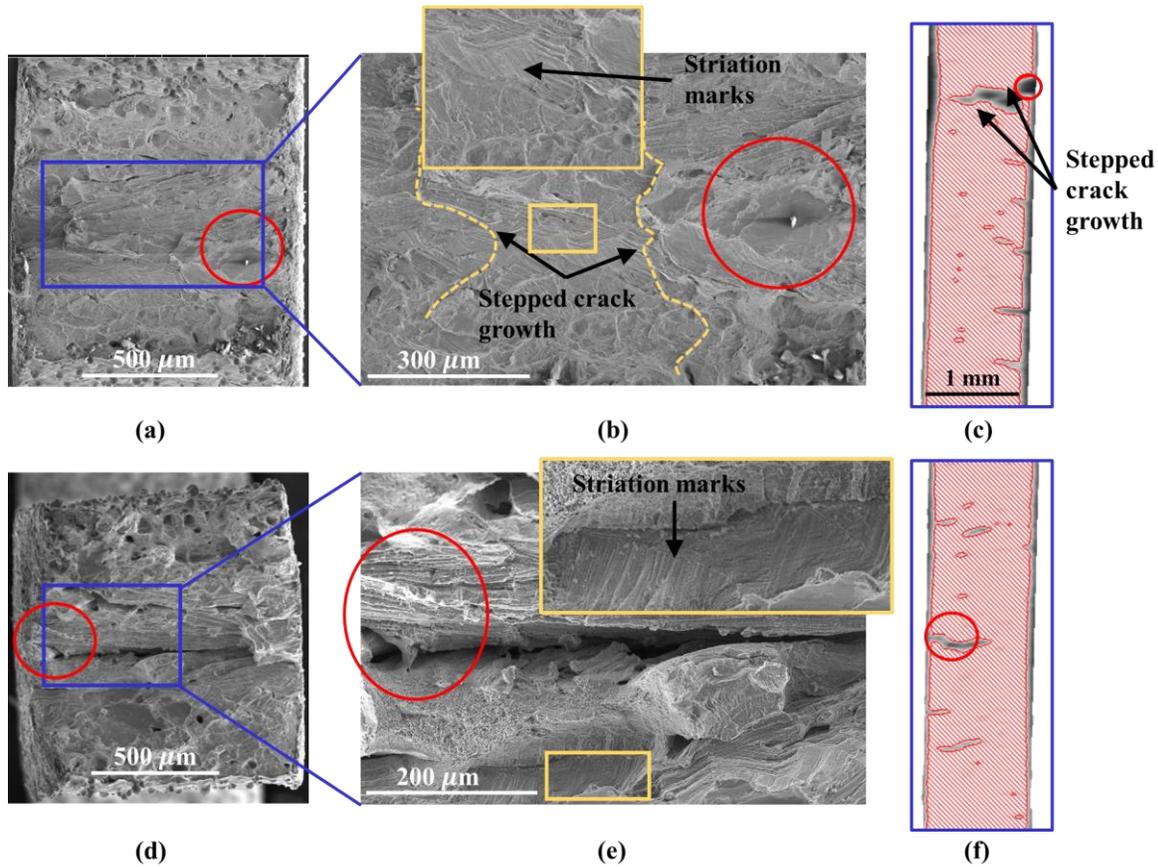

**Figure 13. Representative fractography image of the S3 SS316L specimen: (a) the crack initiation zone and (b) a zoomed-in image with striation marks and stepped crack growth. (c) The crack initiation site is correlated with the cross-sectional CT image. Representative fractography image of S5 SS316L specimen: (d) the crack initiation zone and (e) zoomed-in image with striation marks. (f) The crack initiation site is corroborated by the cross-sectional CT image. The crack initiation zones are demarcated by the red ellipse on the fractured surfaces.**

### 3.5 Functional relationship between initiation life, fatigue life, and strain amplitude

From the experimental observations, fatigue and initiation life are described as functions of material properties and strain amplitude. The Coffin-Manson-Basquin equation provides an approximate fatigue life of a particular material, and an empirical relationship between initiation life and fatigue life can predict the damage initiation instance in a component. The knowledge of the initiation life of a material system facilitates the engineers in recommending optimum operational conditions for a component [82]. Furthermore, cycles to initiation and cycles to failure inform about the crack growth rate in that material system [83]. Therefore, the objective of design engineers is to decrease the crack growth rate in a component.

To formulate the functional relationship between initiation and fatigue life, the fatigue lives of AlSi10Mg specimens are plotted against the corresponding strain amplitudes, and a curve is fitted to the experimental observations with a determination coefficient ($R^2$) value of 0.92, as shown in Figure 14(a). The Coffin-Manson-Basquin parameters for equation (9) are extracted from the curve as presented in Table 3, and the parameters are validated with the values reported in the open literature [84]. The agreement with the literature confirms that the testing apparatus provides reliable experimental observations, facilitating the formulation of a useful correlation. In Figure 14(b), initiation life from the compliance technique is plotted against fatigue life, and an empirical relationship (equation (10)) is derived from the fitted curve with an $R^2$ value of 0.99. The fatigue life from equation (10) is substituted in equation (9) to establish a relationship between strain amplitude and initiation life of AlSi10Mg specimens as shown in equation (11).

$$\frac{\Delta\epsilon}{2} = \left(\frac{\sigma_f'}{E}\right)\left(2N_f\right)^b + \epsilon_f'\left(2N_f\right)^c \tag{9}$$

$$N_i = 0.75 N_f^{1.015} \tag{10}$$

Here, $\Delta\epsilon/2$ is the strain amplitude, $\sigma_f'$ is the fatigue strength coefficient, E is the elastic modulus, $b$ is the fatigue strength exponent, $\epsilon_f'$ is the fatigue ductility coefficient, $c$ is the fatigue ductility exponent, $N_i$ is the initiation life, and $N_f$ is the fatigue life. Substituting $N_f$ from Equation (10) in Equation (9),

$$\frac{\Delta\epsilon}{2} = \left(\frac{\sigma_f'}{E}\right)\left(\frac{2N_i}{0.75}\right)^{\frac{b}{1.015}} + \epsilon_f'\left(\frac{2N_i}{0.75}\right)^{\frac{c}{1.015}} \tag{11}$$

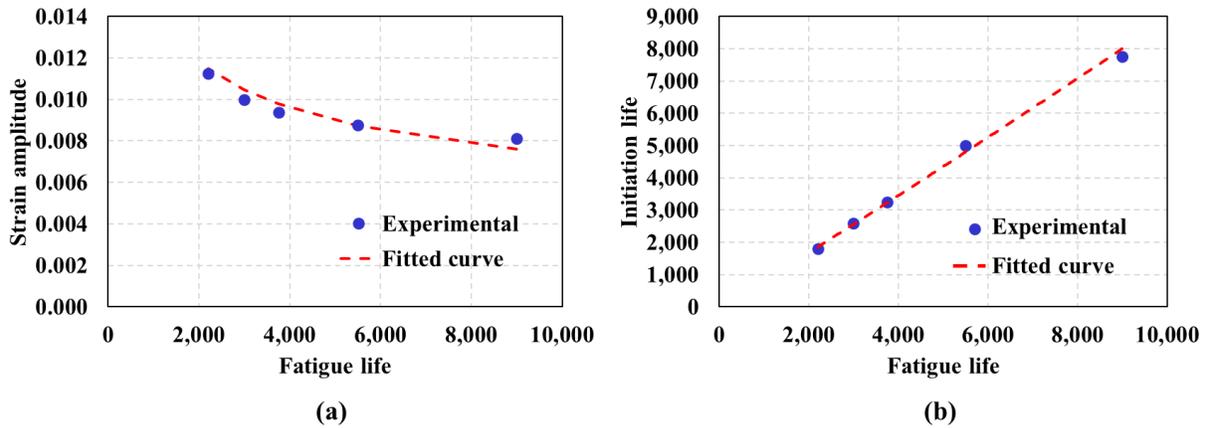

**(a)**

**(b)**

**Figure 14. (a) Variation of AlSi10Mg fatigue life with strain amplitude along with fitted Coffin-Manson-Basquin curve and (b) variation of initiation life with fatigue life along with fitted curve.**

**Table 3. Coffin-Manson-Basquin parameters for AlSi10Mg specimens.**

|  | Fatigue strength coefficient ($\sigma_f'$) (MPa) | Fatigue strength exponent ($b$) | Fatigue ductility coefficient ($\epsilon_f'$) | Fatigue ductility exponent ($c$) |
|---|---|---|---|---|
| Current work | 320 | -0.07 | 0.2 | -0.37 |
| Literature [84] | 320 | -0.07 | 0.1 | -0.56 |

A similar procedure is followed to correlate the initiation life of the SS316L specimens with strain amplitude and material properties. The variation of fatigue life with strain amplitude is plotted in Figure 15(a), and a curve is fitted to the observation with an $R^2$ value of 0.9. From the fitted curve, Coffin-Manson-Basquin parameters are extracted and compared with the existing literature [85], as presented in Table 4. The comparison is followed by the derivation of the correlation between the initiation and fatigue life of SS316L specimens. To derive the correlation, initiation life is plotted against the fatigue life of the specimens and a best-fit curve is generated with an $R^2$ value of 0.99, as presented in Figure 15(b). The derived correlation between initiation and fatigue life is presented in equation (12), which is substituted in the Coffin-Manson-Basquin relation (equation (9)) to formulate a relationship between strain amplitude, material properties, and initiation life (equation (13)). A similar empirical relationship is reported in the open literature for rolled high-strength steel [86].

$$N_i = 0.65 N_f^{1.02} \tag{12}$$

$$\frac{\Delta \epsilon}{2} = \left(\frac{\sigma_f'}{E}\right)\left(\frac{2N_i}{0.65}\right)^{\frac{b}{1.02}} + \epsilon_f'\left(\frac{2N_i}{0.65}\right)^{\frac{c}{1.02}} \tag{13}$$

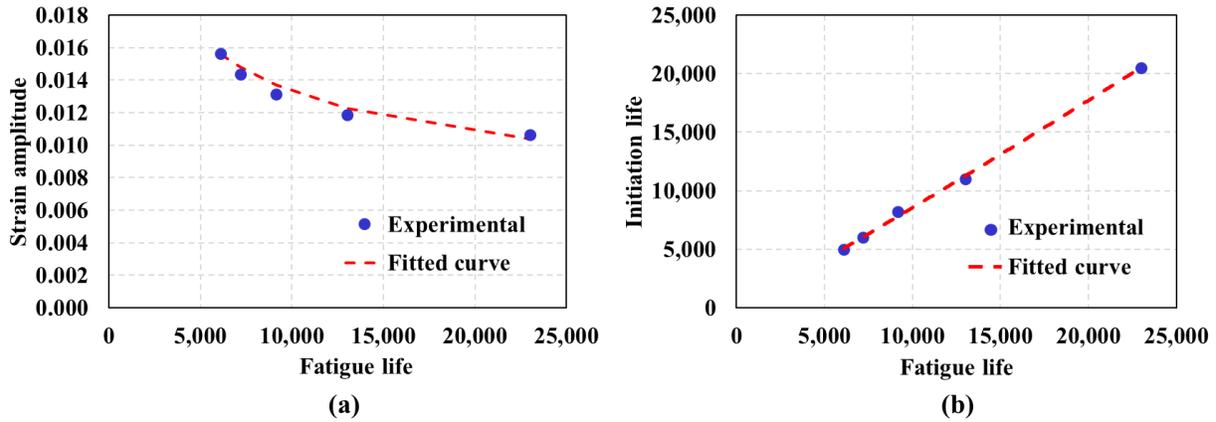

**Figure 15. (a) Variation of SS316L fatigue life with strain amplitude along with fitted Coffin-Manson-Basquin curve and (b) variation of initiation life with fatigue life along with fitted curve.**

**Table 4. Coffin-Manson-Basquin parameters for SS316L specimens.**

|  | Fatigue strength coefficient ($\sigma_f'$) (MPa) | Fatigue strength exponent ($b$) | Fatigue ductility coefficient ($\epsilon_f'$) | Fatigue ductility exponent ($c$) |
|---|---|---|---|---|
| Current work | 1,000 | -0.08 | 0.8 | -0.45 |
| Literature [85] | 1,000 | -0.08 | 0.7 | -0.59 |

The correlations between initiation and fatigue lives of AlSi10Mg suggest that crack initiation occurs around 75% of fatigue life. Similar values were reported in the work of Dharmadhikari et al., which encompassed low cycle fatigue crack initiation investigation of L-PBF AlSi10Mg specimens [47]. On the contrary, cracks in SS316L specimens originate around 65% of fatigue life. These percentage values agree with the values reported in the literature for SS316L specimens [87]. These observations further corroborate the stark difference in the material behavior of the alloys. As discussed in earlier sections, the shorter crack propagation period in AlSi10Mg specimens compared to SS316L specimens can be attributed to the brittle behavior of AlSi10Mg.

## 4. Conclusions

The current work utilizes an analytical initiation life evaluation methodology independent of material systems. The technique encompasses compliance evaluation from the raw load-displacement data. In-situ fatigue crack investigation inside the CT facility is used to correlate the results obtained from the proposed methodology. From the investigations, the following inferences are drawn:

(i) Crack initiation in a specimen is always associated with a load drop across the specimen, which becomes prominent with compliance evaluation.

(ii) The compliance-based methodology produces reliable results for initiation life, which are validated with analytical load drop calculation and correlated with in-situ fatigue crack investigation.

(iii) Crack initiation in the AlSi10Mg specimens is rapidly followed by final fatigue failure without any significant crack propagation zone, unlike SS316L specimens, where the crack propagates throughout the specimen with a negligible rapid failure zone.

(iv) Crack initiation in AlSi10Mg begins at around 75% of the total fatigue life, whereas for SS316L specimens, initiation occurs at around 65% of the fatigue life. This difference can be attributed to the brittle behavior of AlSi10Mg and the ductile behavior of SS316L specimens.

Since the compliance methodology relies solely on load and displacement data, in the future, it will be applied to thermomechanical fatigue experiments. Furthermore, the compliance methodology will be combined with data-driven and numerical models to predict fatigue initiation. Additionally, this methodology will be implemented in complex multi-axial stress conditions.

**Funding information**

The research is funded in part by the Department of Mechanical Engineering (Penn State University) and the Center for Biodevices (Penn State University). Any opinions, findings, and conclusions in this paper are those of the authors and do not necessarily reflect the views of the supporting institution.

**Declaration of Competing Interest**

The authors declare that they have no known competing financial interests or personal relationships that could have appeared to influence the work reported in this paper.

**Acknowledgments**

The authors would like to thank the members of the Centre of Quantitative Imaging Laboratory at Penn State for their help in conducting the in-situ mechanical testing inside the computed tomography facility. The authors would like to thank the Center for Innovative Materials Processing through Direct Digital Deposition (CIMP-3D) at Penn State and Lawrence Livermore National Laboratory for their help in fabricating AlSi10Mg and SS316L specimens, respectively. The authors would like to thank Dr. Catherine Berdanier for her help with scientific writing through the course ME 597: Advanced Engineering Writing at Penn State.

**CRediT authorship contribution statement**

**Ritam Pal:** Investigation, Data curation, Formal analysis, Methodology, Software, Validation, Visualization, Writing - original draft. **Amrita Basak:** Conceptualization, Methodology, Resources, Supervision, Funding acquisition, Project administration, Writing - review & editing. During this work's preparation, OpenAI ChatGPT has been used to improve grammar and readability.